\documentclass[apj]{emulateapj}

\shorttitle{Two-dimensional magnetohydrodynamical case}

\begin{document}

\title{Magneto-hydrodynamical Numerical simulation of wind production from black hole hot accretion flows at very large radii}

\author{De-Fu Bu\altaffilmark{1} and Feng Yuan\altaffilmark{1}, Zhao-Ming Gan\altaffilmark{1}, Xiao-Hong Yang\altaffilmark{2}}

% add the spectrum?

\altaffiltext{1}{Key Laboratory for Research in Galaxies and
Cosmology, Shanghai Astronomical Observatory, Chinese Academy of
Sciences, 80 Nandan Road, Shanghai, 200030, China; fyuan@shao.ac.cn}
\altaffiltext{2}{Department of Physics, Chongqing University,
Chongqing 400044, China}
%\altaffiltext{2}{JILA, University of Colorado and National Institute of Standards and Technology, 440 UCB, Boulder, CO 80309, USA}

\begin{abstract}
Numerical simulations of black hole hot accretion flows have shown
the existence of strong wind.  Those works focus only on the region
close to black hole thus it is unknown whether or where the wind
production stops at large radii. To address this question, Bu et al.
(2016) have performed hydrodynamic (HD) simulations by taking into
account the gravitational potential of both the black hole and the
nuclear star clusters. The latter is assumed to be $\propto \sigma^2
\ln(r)$, with $\sigma$ being the velocity dispersion of stars and
$r$ be the distance from the center of the galaxy. It was found that
when the gravity is dominated by nuclear stars, i.e., outside of
radius $R_A\equiv GM_{\rm BH}/\sigma^2$, winds can no longer be
produced. That work, however, neglects the magnetic field, which is
believed to play a crucial dynamical role in the accretion and thus
must be taken into account. In this paper, we revisit this problem
by performing magneto-hydrodynamical (MHD) simulations. We confirm
the result of Bu et al. (2016), namely wind can't be produced at the
region of $R>R_A$. Our result, combined with the results of Yuan et
al. (2015), indicates that the formula describing the mass flux of
wind $\dot{M}_{\rm wind}=\dot{M}_{\rm BH}(r/20r_s)$ can only be
applied to the region where the black hole potential is dominant.
Here $\dot{M}_{\rm BH}$ is the mass accretion rate at the black hole
horizon and the value of $R_A$ is similar to the Bondi radius.
\end{abstract}

\keywords {accretion, accretion disks $-$ black hole physics $-$
hydrodynamics}

\section{INTRODUCTION}

Hot accretion flows are very common in the universe, ranging from
low-luminosity active galactic nuclei, which is the majority of
nearby galaxies, to the quiescent and hard states of black hole
X-ray binaries (see Yuan \& Narayan 2014 for a review of our current
theoretical understanding of hot accretion flow and its
astrophysical applications). One of the most important progresses in
our understanding of hot accretion flows in recent years is the
finding of strong wind (Yuan et al. 2012b, hereafter YBW12; Narayan
et al. 2012; Li et al. 2013; Yuan et al. 2015; Gu 2015). This is an
important topic since wind is not only an important ingredient of
accretion physics but also plays an important role in AGN feedback
(e.g., Ostriker et al. 2010). Because of the mass loss via wind, the
mass inflow rate decreases inwards $\dot{M}_{\rm {in}}(r)\propto
r^s$ with $s\sim 0.5-1$ (see a short review of various numerical
simulations in Yuan et al. 2012a). This theoretical prediction was
confirmed by the 3 million seconds {\it Chandra} observation to the
supermassive black hole in our Galactic center, Sgr A* (Wang et al.
2013).

The detailed properties of winds such as the mass flux and terminal
velocity are calculated by Yuan et al. (2015). This is achieved by
performing trajectory analysis of virtual test particles based on
three dimensional general relativistic magneto-hydrodynamic (GRMHD)
simulation data. Among others, they find that the mass flux of wind
can be described by $\dot {M}_{\rm {wind}}\approx \dot {M}_{\rm
{BH}}(r/20r_s)$, with $\dot {M}_{\rm {BH}}$ is the mass accretion
rate at the black hole horizon and $r_s$ is the Schwarzschild
radius.  From this equation, most of the wind comes from the region
of large radius. Then a question is how large the value of $r$ can
be. To investigate this question, Bu et al. (2016, hereafter Paper
I) have performed HD simulations to study the accretion flow at very
large radii. We take into account the gravity of both the central
black hole and the nuclear star clusters. The velocity dispersion of
stars is assumed to be a constant and the gravitational potential of
the nuclear star cluster $\propto \sigma^2 \ln(r)$, where $\sigma$
is the velocity dispersion of stars. We find that there is very few
wind launched from the accretion flow in the region where the
gravity is dominated by the star cluster.

In Paper I, we introduce an anomalous stress to transfer the angular
momentum. In reality, it is the MHD turbulence associated with the
magneto-rotational instability (MRI; Balbus \& Hawley 1991, 1998)
that is responsible for the angular momentum transfer. So MHD
simulation is more realistic. Therefore in this paper, we continue
our study by performing MHD simulations. As in Paper I, we define a
radius $R_A$ at which the gravitational force due to the central
black hole is equal to that due to the nuclear star cluster. We call
$R_A$ to be the boundary of the accretion flow or active galactic
nuclei (AGNs). This is partly because the value of $R_A$ is in
reality roughly equal to that of the Bondi radius (Paper I).
Hereafter, we use BHAF to refer the hot accretion flow close to the
black hole. We use CAAF (circum-AGN accretion flow) to refer the hot
accretion flow at larger radii beyond $R_A$.

The structure of the paper is as follows. In \S 2, we describe
the basic equations and the set up of the initial conditions. The
results of simulations will be given in \S 3. We discuss and
summarize our results in \S 4.

\section{method}\label{method}
\subsection{Equations}

In a spherical coordinate $(r, \theta, \phi)$, we solve the following
magnetohydrodynamical equations describing accretion:
\begin{equation}
\frac{d\rho}{dt}+\rho\nabla\cdot \mathbf{v}=0,\label{cont}
\end{equation}
\begin{equation}
\rho\frac{d\mathbf{v}}{dt}=-\nabla p-\rho\nabla
\psi+\frac{1}{4\pi}(\nabla \times \mathbf{B}) \times \mathbf{B},
\label{monentum}
\end{equation}
\begin{equation}
\rho\frac{d(e/\rho)}{dt}=-p\nabla\cdot\mathbf{v}+\eta\mathbf{J}^2,
\label{energyequation}
\end{equation}
\begin{equation}
\frac{\partial \mathbf{B}}{\partial t}=\nabla \times (\mathbf{v}
\times \mathbf{B}-\eta \mathbf{J}). \label{inductionequation}
\end{equation}
Here, $\rho$, $p$, $\mathbf{v}$, $\psi$, $e$, $\mathbf {B}$ and
$\mathbf{J}(=(c/4\pi) \nabla \times \mathbf{B})$ are density,
pressure, velocity , gravitational potential, internal energy,
magnetic field and the current density, respectively. $d/dt(\equiv
\partial / \partial t+ \mathbf{v} \cdot \nabla)$ denotes the Lagrangian
time derivative. We adopt an equation of state of ideal gas
$p=(\gamma -1)e$, and set $\gamma =5/3$.

The gravitational potential $\psi$ can be expressed as
\begin{equation}
\psi= \psi_{BH}+\psi_{star}.
\end{equation}
The black hole potential $\psi_{BH}=-GM_{BH}/(r-r_s)$, where $G$ is
the gravitational constant, $M_{BH}$ is the mass of the black hole
and $r_s$ is the Schwarzschild radius. As in Paper I, in this paper,
we assume that the velocity dispersion of nuclear stars is a
constant of radius. This seems to be the case of many AGNs. So the
potential of the star cluster is $\psi_{star}=\sigma^2 \ln (r)+C$,
where $\sigma$ is the velocity dispersion of stars and $C$ is a
constant. So we have \begin{equation} R_A=GM_{\rm {BH}}/\sigma^2.
\label{boundary}
\end{equation} We set $\sigma^2=10$ and $G=M_{\rm BH}=1$ to define
our units in the present work. Under this units we have $R_A=0.1$.
For a  typical physical value of $\sigma\sim (100-400) {\rm
km~s^{-1}}$ (e.g., Kormendy \& Ho 2013), $R_A \sim (10^5-10^6) r_s$.
Figure 1 in Paper I shows the gravitational force distribution.

In the above equations, the final terms in Equations
(\ref{energyequation}) and (\ref{inductionequation}) are the
magnetic heating and dissipation rate mediated by a finite
resistivity $\eta$. The exact form of $\eta$ is same as that used in
Stone \& Pringle (2001). In the code we adopt, the energy equation
is an internal energy equation, numerical reconnection inevitably
results in loss of energy from the system. By adding the anomalous
resistivity $\eta$, the energy loss can be captured in the form of
heating in the current sheet (Stone \& Pringle 2001).

In this paper, time is expressed in unit of the orbital time at the
torus center.

\subsection{Initial conditions}
As for the initial condition, we assume a rotating equilibrium torus
embedded in a non-rotating, low-density medium. We assume that the
torus has constant specific angular momentum $L$ and assume a
polytropic equation of state, $p=A\rho^\gamma$, where $A$ is a
constant. The density distribution of torus is \begin{equation}
\rho=\rho_c \left \{\frac{\max[\Psi(R_0,\pi/2)-\psi(r,
\theta)-L^2/(2(r
\sin\theta)^2),0]}{A[\gamma/(\gamma-1)]}\right\}^{1/(\gamma-1)},
\end{equation} where $R_0$ is the density
maximum (center) of the torus (Nishikori et al. 2006), $\rho_c$ is
the density at the torus center. In this paper, we assume $\rho_c=1$
and $A=0.4$.

The ambient medium in which the torus is embedded has density
$\rho_0$ and pressure $\rho_0/r$. The mass and pressure of the
ambient medius are negligibly small, we choose $\rho_0=10^{-4}$.

\subsection{Models}
%In this paper, we both simulate CAAFs and BHAF. In models A1-A4
%which simulate the CAAFs,
The initial magnetic field is generated by a vector potential, i.e.
$\mathbf{B}=\nabla \times \mathbf{A}$. In models A1 and A2, the
initial magnetic field has a dipolar configuration (same as that in
Stone \& Pringle 2001). We take $\mathbf{A}$ to be purely azimuthal
with $A_\phi=\rho^2/\beta$, with $\beta=200$. The only difference
between models A1 and A2 is that the resolution of model A1 is two
times of that of model A2. We find that, the results for models A1
and A2 are almost same. So the resolution in model A2 is enough for
our problem. In order to study the dependence of results on initial
magnetic field configuration, we carry out model A3. In this model,
the initial magnetic field has a quadrapolar configuration with
$A_\phi=\rho^2/\beta r \cos \theta$, and $\beta=100$. Table 1
summarizes the models.

For the initial condition in models A1, A3 and A4, over most of the
central regions of the initial torus, we have $6$ grids for one
wavelength of the fastest growing mode. Therefore, the fastest
growing model of MRI is marginally resolved in our simulations.

In this paper, the simulations are two-dimensional (hereafter 2D).
According to the antidynamo theorem (Cowling 1933; Sadowski et al.
2015), the turbulence induced by MRI can not be self-sustaining.
Therefore there can not be a true steady state and the quasi-steady
state of the simulations is only transient. In this paper, we still
preform 2D simulation because on one hand we can simulate a larger
radial dynamical range, and on the other hand previous works have
indicated that for many problems the results from 2D and 3D
simulations are often quite similar (e.g., see the short review in
Yuan et al. 2012a in the case of radial profile of inflow and
outflow rates). Still, for our present study of wind from accretion
flows, it is necessary to carefully examine whether the results from
2D simulation are consistent with those from 3D simulation.

In order to answer this question, we have carried out a 2D MHD
simulation of accretion flows close to a black hole. The domain of
the simulation is $2r_s$-$400r_s$. In this simulation, only the
black hole gravity is taken into account. Using the trajectory
approach as described in Yuan et al. (2015), we have calculated the
mass flux of wind, which is $\sim 52\%$ of the total outflow rate
calculated by Equation (\ref{outflowrate}). As comparison, the mass
flux of wind calculated in Yuan et al. (2015), which is based on 3D
MHD simulation data of accretion flow, is $\sim 60\%$ of the total
outflow rate calculated by Equation (\ref{outflowrate}). Such a good
consistency suggests that the present 2D study should be a good
approximation.

\begin{table}
\footnotesize
\begin{center}
\caption{models in this paper}
\begin{tabular}{cccc} \\ \hline
Models & Initial magnetic field & Resolution  & Computational domain \\ %&  Computational domain  & Gravitational potential\\
\hline

A1 &  dipolar        & $294 \times 160$ & 0.03-4 \\ %& $0.02 \leq r \leq 4$ & Star cluster gravity + Black hole gravity\\
A2 &  dipolar        & $147 \times 80 $ & 0.03-4 \\ %& $0.002 \leq r \leq 0.4$ & Star cluster gravity + Black hole gravity\\
A3 &  quadrapolar    & $294 \times 160$ & 0.02-4\\ %& $0.02 \leq r \leq 4$ & Star cluster gravity + Black hole gravity\\
A4 &  dipolar        & $294 \times 160$ & 0.002-0.4\\
%B  &  $ \nu \propto r^{1/2}$    & $334 \times 160$ \\ %& $2r_s \leq r \leq 400r_s$ &   Black hole gravity\\
\hline
\end{tabular}\\
\end{center}
\end{table}

\subsection{Numerical Method}
We use the ZEUS-2D code (Stone \& Norman 1992a, 1992b) to solve
Equations (1)-(4). The polar range is $0 \leq \theta \leq \pi$. We
adopt non-uniform grid in the radial direction $(\bigtriangleup
r)_{i+1} / (\bigtriangleup r)_{i} = 1.037$. The distributions of
grids in $\theta$ direction in the northern and southern hemispheres
are symmetric about the equatorial plane. The resolution at $\theta$
in the northern hemisphere is same as that at $\pi-\theta$ in the
southern hemisphere. In order to well resolve the accretion disk
around the equatorial plane, the resolution is increased from the
north and south rotational axis to the equatorial plane with
$(\bigtriangleup \theta)_{j+1} / (\bigtriangleup \theta)_{j} =
0.9826$ for $0 \leq \theta \leq \pi/2$ and $(\bigtriangleup
\theta)_{j+1} / (\bigtriangleup \theta)_{j} = 1.0177$ for $\pi/2
\leq \theta \leq \pi$.  At the poles, we use axisymmetric boundary
conditions. At the inner and outer radial boundary, we use outflow
boundary conditions.

\section{RESULTS}\label{results}

\subsection{Mass inflow rate}
Following Stone et al. (1999), we define the mass inflow and outflow
rates, $\dot {M}_{in}$ and $\dot {M}_{out}$ as follows,

\begin{equation}
 \dot{M}_{\rm in}(r) = 2\pi r^{2} \int_{0}^{\pi} \rho \min(v_{r},0)
   \sin \theta d\theta,
   \label{inflowrate}
\end{equation}
\begin{equation}
 \dot{M}_{\rm out}(r) = 2\pi r^{2} \int_{0}^{\pi} \rho \max(v_{r},0)
    \sin \theta d\theta.
    \label{outflowrate}
\end{equation}
The net mass accretion rate is,
\begin{equation}
\dot{M}_{\rm acc}(r)=\dot{M}_{\rm in}(r)+\dot{M}_{\rm out}(r).
\label{netrate}\end{equation} Note that the above rates are obtained
by time-averaging the integrals rather than integrating the time
averages.

\begin{figure}
\includegraphics[width=8cm]{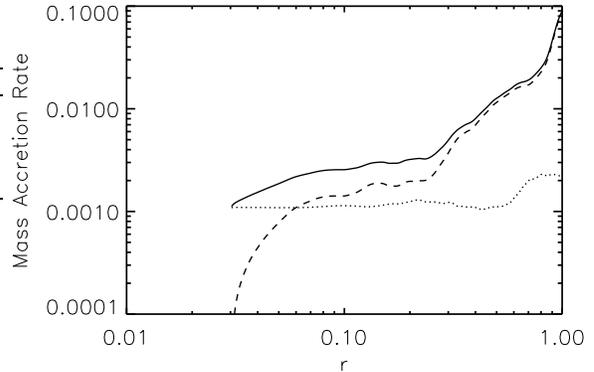}
\caption{The radial profile of the time-averaged (from $t=130$ to
136 orbits) and angle integrated mass inflow rate $\dot{M}_{\rm in}$
(solid line), outflow rate $\dot{M}_{\rm out}$ (dashed line), and
the net rate $\dot{M}_{\rm acc}$ (dotted line) in model A1. They are
defined in Equations (\ref{inflowrate}), (\ref{outflowrate}), and
(\ref{netrate}), respectively. \label{Fig:accretionA1}}
\end{figure}

Figure \ref{Fig:accretionA1} shows the time-averaged (from 130-136
orbits) and angle-integrated mass rates of model A1. Both the mass
inflow and outflow rates decrease inward. This is consistent with
that found in the HD simulations in Paper I. We note that the mass
inflow and outflow rates are not good power-law function of radius,
just like the case of a BHAF (Stone \& Pringle 2001). This is likely
because the radial distribution of the strength of magnetic field is
not very smooth. Because of the accumulation of magnetic flux during
the simulation, the magnetic field in the inner region of the
accretion flow is much stronger than the other region. We will see
in \S3.4 that, in model A3 there is no such accumulation of magnetic
flux due to the initial quadrapolar configuration, the radial
profiles become smoother.

To reach a steady state at a certain radius, the simulation time
should be at least equal to the accretion timescale at that radius.
According to this criterion, our simulation of model A1 has reached
a steady state within $r\sim 0.6$. We note that the flow in model A1
is convectively unstable (see section 3.3). The accretion timescale
is roughly equal to one turnover time of local convective eddies. It
may require many turnover times for the convection to reach a steady
state. Thus it will be interesting to run our simulations for
several times longer in the future to check whether the results will
change.

\begin{figure}
\includegraphics[width=13cm]{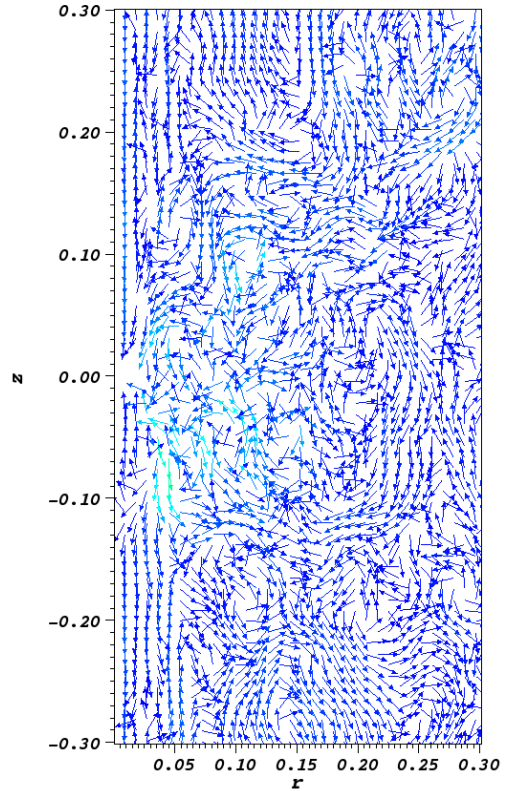}
\vspace{-1cm} \caption{Snapshot of velocity vector of model A1 at
$t=130$ orbits. Clearly, turbulent eddies occupy the whole domain
and it is hard to find winds.  \label{Fig:vectorA1}}
\end{figure}

\begin{figure}
\includegraphics[width=8cm]{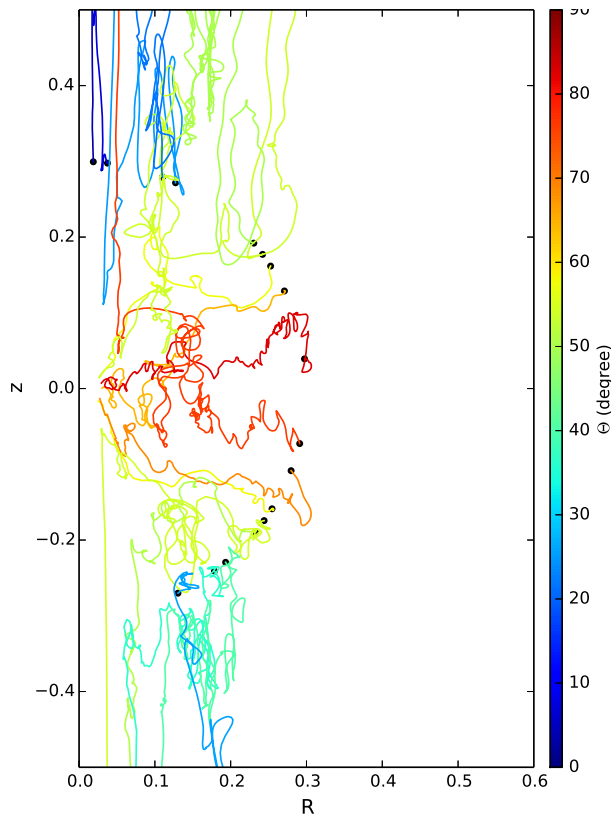}
\caption{Trajectory of gas for model A1. The black dots located at
$r=0.3$ are starting points of the ``test particle''. Different
colors denote trajectory of ``test particle'' starting from
different $\theta$ angle. It is clear that the ``test particle''
crosses the starting radius for many times. From this figure we can
see that the real wind trajectories, i.e., the trajectories which
extend from $r=0.3$ to large radius and never come across $r=0.3$
again are very few. Winds are very weak. \label{Fig:trajectoryA1}}
\end{figure}

\subsection{Does strong wind exist in a CAAF ?}

The significant mass outflow rate shown in Figure
\ref{Fig:accretionA1} does not mean the existence of strong  real
outflow (wind) because it may be due to the turbulent motion. To
study whether wind exists, let us first directly look at the
velocity field shown in Figure \ref{Fig:vectorA1}. We see that
turbulent eddies occupy the whole domain and it is hard to find
systematic winds.

To investigate this problem precisely, following Yuan et al. (2015),
we use the trajectory method to study the motion of the virtual
particles. The details of this approach can be found in Yuan et al.
(2015).  To get the trajectory, we first need to choose some virtual
``test particle'' in the simulation domain. They are of course not
real particles, but some grids representing fluid elements. Their
locations and velocities at a certain time $t$ are obtained directly
from the simulation data. We can then obtain their location at time
$t +\delta ¦Ät$ from the velocity vector and $\delta t$. Trajectory
is more loyal than streamline to reflect the motion of particles
which is crucial for us to investigate whether real wind exists.
Trajectory is only equivalent to the streamline for strictly steady
motion, which is not the case for accretion flow since it is always
turbulent.

Trajectory approach can easily tell us which particles are real
outflows (i.e., winds) and which are doing turbulent motions.
Combining with the density and velocity field information from the
simulation data we can then obtain the various properties of wind
such as the mass flux, angular distribution, and velocity (see Yuan
et al. 2015 for details). Figure \ref{Fig:trajectoryA1} shows the
trajectory of some gas particles starting from $r=0.3$ in model A1.
From this figure we can see that the real wind trajectories, i.e.,
the trajectories which extend from $r=0.3$ to large radius and never
come across $r=0.3$ again, are very few. This implies that the mass
flux of wind is very small. Our quantitative calculation confirms
this result. For example, we find at $r=0.4$ the ratio of mass flux
of winds to the total outflow rate calculated by Equation
(\ref{outflowrate}) is only $0.2\%$. This result means that there is
almost no wind. This result is consistent with that found in Paper
I. As a comparison, in the case of BHAF, Yuan et al. (2015) find
that this ratio is $60\%$.

\subsection{Why the inflow rate decreases inward in a CAAF?}

If wind is absent, what is the reason for the inward decrease of
inflow rate? To answer this question, we need to examine the
convective stability of the accretion flow. Hydrodynamical
simulation of BHAFs (e.g., Stone et al. 1999; Igumenshchev \&
Abramowicz 1999, 2000; Yuan \& Bu 2010) have found that the flows
are convectively unstable, consistent with what has been suggested
by the one-dimensional analytical study of BHAFs (Narayan \& Yi
1994). The physical reason is that the entropy of the flow increases
inward, which is resulted by the viscous heating and negligible
radiative loss. However, in the presence of magnetic field,
numerical simulations have found that a BHAF becomes convectively
stable (Narayan et al. 2012; YBW12). In the case of a CAAF, Paper I
has found that the flow is convectively unstable, same as the case
of a BHAF.

We now study whether a CAAF is convectively stable or not in the presence of magnetic field.
We use the H{\o}iland criteria (e.g., Tassoul 1978; Begelman \& Meier 1982):
\begin {equation}(\nabla s \cdot \mathbf{dr})(\mathbf{g} \cdot
\mathbf{dr})-\frac{2\gamma v_{\phi}}{R^2}[\nabla(v_{\phi}R)\cdot
\mathbf{dr}]dR <0.\label{hoiland}\end{equation}  In Equation
(\ref{hoiland}), $R=r\sin \theta$ is the cylindrical radius, ${\bf
dr} = dr \hat r + r d \theta \hat \theta$ is the displacement
vector, $s = \ln(p) - \gamma \ln(\rho)$ is $(\gamma - 1)$ times the
entropy, ${\bf g} = - \nabla \psi + \hat R v_{\phi}^2/R$ is the
effective gravity, and $v_{\phi}$ is the rotational velocities. For
a non-rotating flow, this condition is equivalent to an inward
increase of entropy, which is the well-known Schwarzschild criteria.

\begin{figure}\hspace{-1.0cm}
\epsscale{1.2} \plotone{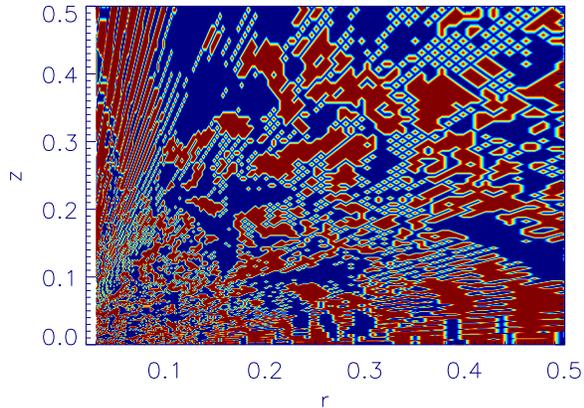}\hspace{-1.cm} \epsscale{1.5}
\caption{Convective stability analysis of Model A1. The result is
obtained according to Equation (\ref{hoiland}) based on the
simulation data at t=132 orbits at the initial torus center $r=1$.
The red region is unstable.} \label{Fig:instability}
\end{figure}

Taking model A1 as an example, Figure \ref{Fig:instability} shows
the result. The result is obtained according to Equation
(\ref{hoiland}) based on the simulation data at t=132 orbits at the
initial torus center $r=1$. At $t=132$ orbits, the flow has achieved
a steady state since the net accretion rate averaged between $t=130$
to 136 orbits is a constant of radius (see figure
\ref{Fig:accretionA1}, the dotted line). The red regions are
convectively unstable. We can see from the figure that  a CAAF is
mostly convectively unstable. This is different from the case of a
BHAF with magnetic field. The reason should be due to the change of
the gravitational potential but the detail remains unclear. This
result strongly implies that the inward decrease of inflow rate is
because of convection and it reminds us the scenario of
convection-dominated accretion flow (CDAF) proposed by Narayan et
al. (2000) and Quauaert \& Gruzinov (2000), although that model was
proposed to explain the dynamics of BHAFs rather than CAAFs.

\begin{figure}\hspace{-1.0cm}
\epsscale{1.2} \plotone{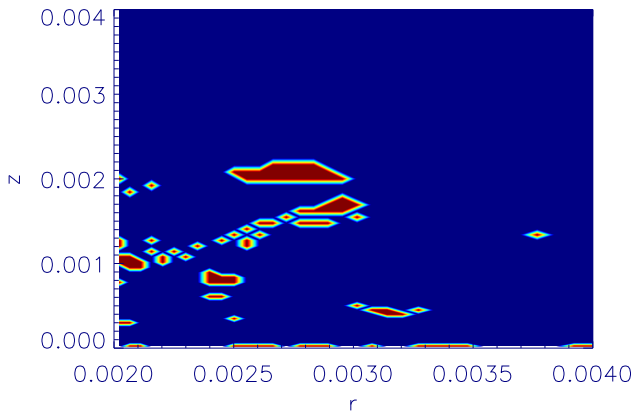}\hspace{-1.cm} \epsscale{1.5}
\caption{Convective stability analysis of Model A4. The result is
obtained according to Equation (\ref{hoiland}) based on the
simulation data at t=100 orbits at the initial torus center $r=0.1$.
The red region is unstable.} \label{Fig:instabilityA4}
\end{figure}

From Figure \ref{Fig:instability}, it seems that the region $r<0.1$
is convectively unstable. Our previous work with only black hole
gravity does show that the flow is convectively stable (YBW12). We
think the apparent discrepancy is because of the contamination of
the gravitational potential by the star cluster. In model A1, in the
region $r<0.1$, it is true that the black hole gravity is bigger
than that of the star cluster, but the gravity of the stars is not
negligible. In fact, in the region $0.03<r<0.1$, the black hole
gravity is bigger than that of star cluster at most by a factor of
4. To further investigate this point, we have analyzed the
convective stability of model A4. Figure \ref{Fig:instabilityA4}
shows the results of the region very close to the black hole where
the black hole gravity is strongly dominant. We can clearly see that
the flow is convectively stable.

\subsection{Varying the initial configuration of magnetic field}
\begin{figure}
\includegraphics[width=8cm]{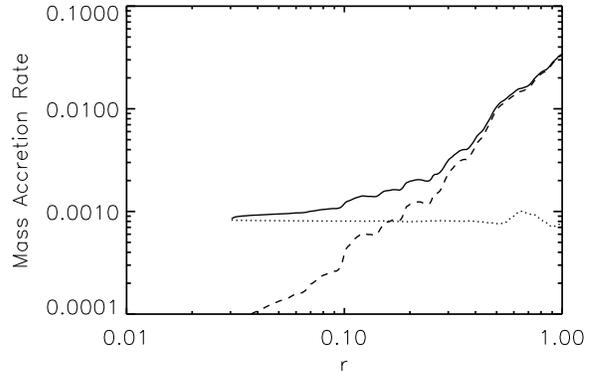}
\caption{Same as Figure \ref{Fig:accretionA1}, but for model A3. The
time-average is taken from $t=127$ to 130 orbits.
\label{Fig:accretionA3}}
\end{figure}

In order to study whether the results depend on the initial
configuration of the magnetic field, we carry out model A3. In this
model, the initial configuration of magnetic field is quadrapolar.
Figure \ref{Fig:accretionA3} shows the inflow, outflow, and net
rates. The inflow rate is smoother than that in Figure 1, as we have
explained in \S3.1. In addition to that, we find that all results
are almost same with  those of model A1, namely the flow is
convectively unstable, and wind is absent. Quantitatively, using the
trajectory method, we find at $r=0.4$ , the ratio of mass flux of
winds to the total outflow rate calculated by Equation
(\ref{outflowrate}) is $0.1\%$.

\subsection{Moving the computational domain inward}
\begin{figure}
\includegraphics[width=8cm]{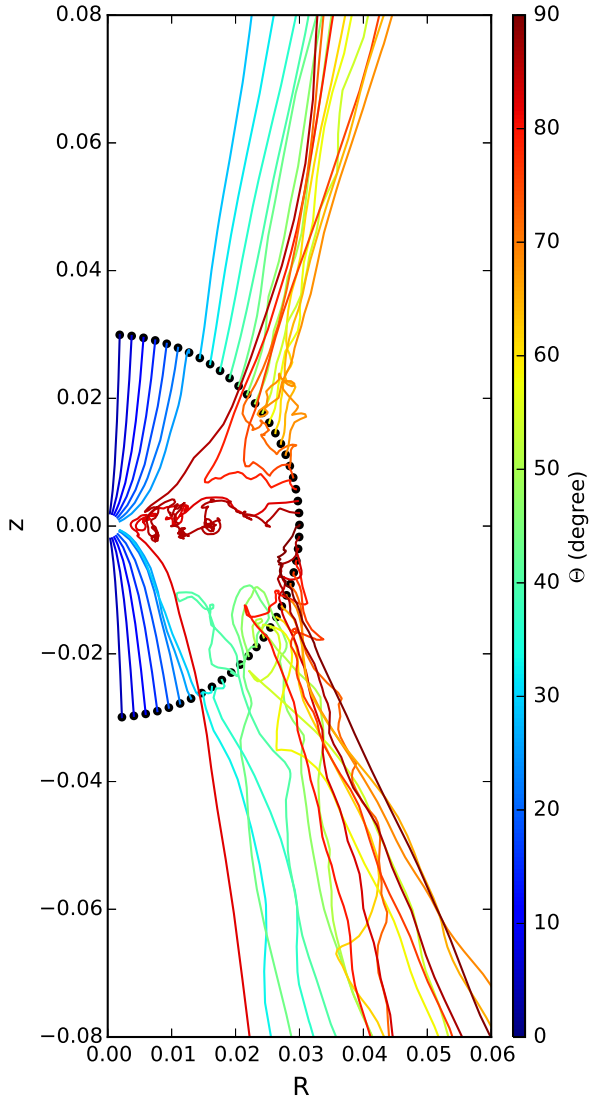}
\caption{Trajectory of gas for model A4. The black dots located at
$r=0.03$ are starting points of the ``test particle''. Different
colors denote trajectory of ``test particle'' starting from
different $\theta$ angle. It is clear that wind is present when the
black gravity dominates. \label{Fig:trajectoryA4}}
\end{figure}
From Figures \ref{Fig:accretionA1} and \ref{Fig:accretionA3}, in the
region $r<0.1$, the outflow rate is very small. This region is
dominated by the gravity of black hole. Previous works (Yuan et al.
2012b; Yuan et al. 2015) have shown that in this case outflow (wind)
should be strong. The apparent discrepancy between this work and our
previous works is due to the fact that the region $r<0.1$ is too
close to the inner boundary where a somewhat ``unphysical'' boundary
condition (i.e., outflow condition) is adopted. This condition means
that the gas entering into the inner boundary is assumed to
disappear and the gradient of physical variables at the boundary is
zero. However, in reality, there should be some flow entering into
the inner boundary and the flows inside and outside of the boundary
can interact with each other. This will significantly affect the
properties of the flows around the inner boundary.

In order to illustrate this point, we have carried out model A4. In
this model, our computational domain is $0.002< r < 0.4$. The
gravitational potential from both the black hole and the nuclear
star cluster are included but obviously the former dominates. The
initial condition of the magnetic field is dipolar. Figure
\ref{Fig:trajectoryA4} shows the trajectory of some gas particles
starting from $r=0.03$. From this figure, it is clear that strong
wind is present in this case. Combined with the results presented in
previous sections, this result indicates that the disappearance of
wind is because of the changes of the gravitational potential.

In models A1 and A4, the setup of the model, the equations, and the
potential formula we use are exactly same. The only difference
between them is that the black hole potential is more dominates in
A4 while the stellar cluster potential is more dominates in A1.
Therefore, the disappearance of wind is because of the changes of
the gravitational potential.

\section{CONCLUSION AND DISCUSSION}\label{sec:summary}
Numerical simulations show that strong winds exist in black hole hot
accretion flow (e.g., YBW12). The mass flux of wind follows
$\dot{M}_{\rm wind}=\dot{M}_{\rm BH}(r/20r_s)$ (Yuan et al. 2015). A
question is then what the value of $r$ can be, i.e., whether or
where the wind production stops. In order to answer this question,
in Paper I, we have performed HD simulations and take into account
the gravity of both the black hole and nuclear stars cluster. We
find that the mass inflow rate decreases inward. However, our
trajectory analysis indicates that there is no wind when the
potential of star cluster dominates, i.e., beyond a certain radius
$R_A\equiv GM_{\rm BH}/\sigma^2$, with $\sigma$ being the velocity
dispersion of stars. The inward decrease of inflow rate is not
because of strong wind, as in the case of accretion flow in which
the black hole potential dominates, but because of the convective
instability of the accretion flow. In this paper, we revisit the
same problem by performing more realistic MHD simulations. We find
again that the conclusion remains unchanged, i.e., there is no wind
beyond $R_A$. Our stability analysis again indicates that the MHD
accretion flow beyond $R_A$ is convectively unstable.  This is
different from the case of accretion flow when the black hole
potential dominates (YBW12; Narayan et al. 2012). So the inward
decrease of inflow rate is likely because of the convective motion
of the flow.

This result indicates that the mass flux of wind found by Yuan et
al. (2015): \begin{equation}\dot{M}_{\rm wind}=\dot{M}_{\rm
BH}(r/20r_s), \end{equation} can only be applied to the region where
the black hole gravitational force dominates. In the star cluster
potential dominates region, i.e., beyond $R_A$,  no wind will be
produced. In practice, the value of $R_A$ is close to the Bondi
radius $R_B\equiv GM_{\rm BH}/c_s^2$ (Paper I).

What is the reason for the absence of wind beyond $R_A$? We
speculate that it may be related with the change of the slope of the
gravitational potential. Such a change  would change the shear of
the accretion flow and then the turbulent stress in the accretion
flow. Analytical models of accretion disk (e.g., Shakura \& Sunyaev
1973) usually assume that viscous stress is proportional to the
shear of the accretion disk, \begin{equation} T_{r\phi}=\rho \nu r
\frac{d\Omega}{dr}  \label{stress} \end{equation} with $\nu=\alpha
c_s^2/\Omega_k$. $c_s$, $\Omega$ and $\Omega_k$ are sound speed,
angular velocity and Keplerian angular velocity, respectively.
Recent 3D MHD numerical simulations show that the turbulent stress
is not linearly proportional to the shear, but the dependence is
stronger than that predicted by Equation (\ref{stress}) (Pessah et
al. 2008; Penna et al. 2013). This implies that the change of the
potential changes some properties of turbulence which then change
the wind production.

\section*{Acknowledgements}
We thank the referee for rasing very useful questions which improve
the paper significantly. We thank Ramesh Narayan for helpful
discussions. This work was supported in part by the National Basic
Research Program of China (973 Program, grant 2014CB845800), the
Strategic Priority Research Program The Emergence of Cosmological
Structures¡± of the Chinese Academy of Sciences (grant XDB09000000)
and the Natural Science Foundation of China (grants 11103061,
11133005, 11121062 and 11573051). This work has made use of the High
Performance Computing Resource in the Core Facility for Advanced
Research Computing at Shanghai Astronomical Observatory.

\label{lastpage}

\end{document}